# ZnO-Based Polariton Laser Operating at Room Temperature: From Excitonic to Photonic Condensate

Feng Li[1,2], Laurent Orosz[3], Olfa Kamoun[4], Sophie Bouchoule[5], Christelle Brimont[4] Pierre Disseix[3], Thierry Guillet[4], Xavier Lafosse[5], Mathieu Leroux[1], Joel Leymarie[3], Meletis Mexis[4], Martine Mihailovic[3], François Réveret[3], Dmitry Solnyshkov[3], Jesus Zuniga-Perez[1], Guillaume Malpuech[3]

[1]CRHEA-CNRS, Rue Bernard Gregory, 06560 Valbonne, FRANCE

[2]Université de Nice Sophia-Antipolis, 06103 Nice, France

[3]Institut Pascal, PHOTON-N2, Clermont Université, CNRS and Université Blaise Pascal, 24 Avenue des Landais, 63177 Aubière cedex, FRANCE

[4]Université de Montpellier 2, CNRS, Laboratoire Charles Coulomb, UMR 5221, 34095 Montpellier, FRANCE

[5]LPN-CNRS, Route de Nozay, 91460 Marcoussis, FRANCE

**A laser threshold is determined by the gain condition, which has been progressively reduced by the use of heterostructures and of quantum confinement[1]. The polariton laser [2,3] is the ultimate step of this evolution: coherent emission is obtained from the spontaneous decay of an exciton-polariton condensate [4,5,6,7,8] without the achievement of any gain condition. ZnO, with its unique excitonic properties [9,10,11,12,13], is the best choice for a blue/UV-emitting polariton laser device[14,15,16,17,18]. We report on the fabrication of a new family of fully hybrid microcavities that combine the best-quality ZnO material available (bulk substrate) and two dielectric distributed Bragg reflectors, demonstrating large quality factors (>2500) and Rabi splittings (~200 meV). Low threshold polariton lasing is achieved between 4 and 300 K and for excitonic fractions ranging between 12% and 96 %. A phase diagram highlighting the role of LO phonon-assisted relaxation in this polar semiconductor is established, and a remarkable switching between polariton modes is demonstrated.**

The unique optical properties of ZnO make it promising for a new generation of blue and UV-emitting diodes and lasers, which are widely used in photonics, information storage, biology, and medicine. In ZnO, the exciton binding energy largely exceeds $k_B T$ at room temperature. The lasing threshold can therefore be lowered by the excitonic emission enhancement. However, ZnO material is technologically difficult to embed in complex photonic structures and ZnO based lasers are typically random, or based on nano-objects with very poor usability of the beams[9,10,11,12]. Photonic confinement, such as in a planar microcavity, allows establishing the strong coupling of excitons with photons, creating exciton-polariton modes[3]. In ZnO, the large exciton oscillator strength can induce an energy splitting between the polariton eigenmodes of up to 200-300 meV, a record for inorganic structures[18,19]. The photon emission time for excitons becomes as short as 20 fs, which is two orders of magnitude faster than in the weak coupling regime. Besides, the absorption of photons in the strong coupling regime is a coherent and reversible process unlike in the weak coupling regime, where it is an irreversible dissipative process. This explains why the gain condition is fully relaxed in a polariton laser. Polariton lasing is therefore only controlled by a balance between the polariton (cavity) life time and the scattering rate towards the condensed state. Depending on the ratio between these two quantities, the condensation can be well described by the thermodynamic theory of Bose Einstein condensation, or be completely out of equilibrium[20,21]. Polariton lasing has been demonstrated at low temperature in CdTe[4] and GaAs[5,22] based microcavities, and the study of fundamental properties of polariton condensates has become a very popular topic in the last years [23,24,25,26].

Room temperature operation requires the use of active materials with large exciton binding energy and oscillator strength. This is why attention has turned to wide-bandgap inorganic semiconductors, such as GaN and ZnO, and to organic materials[27]. Room temperature polariton lasing was demonstrated under optical pumping in GaN-based epitaxial structures combining epitaxial and dielectric Distributed Bragg reflectors (DBRs)[6,7,8]. Following the GaN approach, ZnO-based planar cavities have been fabricated with a bottom epitaxial mirror, an epitaxial ZnO active layer, and a final top dielectric DBR[13,16,28,29]. Vertical Cavity Surface Emitting Lasers[13] and polariton lasers[16,28] were fabricated from such structures, despite a rather low quality factor (200-600) and strong spatial inhomogeneities typical for nitride DBRs. Alternatively, polycrystalline ZnO has been deposited between two dielectric DBRs, thereby sacrificing the crystalline quality of the active material but increasing the cavity quality factor up to 1000, which has also allowed the observation of polariton lasing up to 250 K[17].

In this letter, we report on a new fabrication approach that merges the excellent crystalline properties of bulk ZnO substrates (i.e. defect-free material) with the large quality factors of dielectric DBRs. Polariton laser operation is observed in an unprecedented range of parameters: from 4 to 300 K, and for excitonic fractions of the polariton condensate between 12% and 96 %.

Figure 1 shows a schematic representation of the fabrication steps and of the final structure. First, a $HfO_2/SiO_2$ DBR is deposited on top of the ZnO substrate (fig 1-a). The whole structure is then turned upside down and transferred onto a host substrate[30] (fig 1-b). Next, the active medium is obtained by polishing/etching the 500 µm thick ZnO substrate down to a few hundreds nm (fig 1-c). An important aspect of the process is that neither the optical quality of the remaining ZnO (as shown later) nor its surface are degraded by the previous process. Indeed, the roughness of the ZnO surface after the etching step is typically 5-10 Å in 5×5µm$^2$ images. Finally, the half-cavity is completed by a second HfO$_2$/SiO$_2$ DBR (fig 1-d) whose surface roughness reproduces that of the polished ZnO. The Q factor of the complete cavity deduced from micro-photoluminescence (µPL) experiments is close to 2500 (See Supplementary A). As depicted in figure 1-d, the thickness gradient of the cavity is quasi-continuous. Below threshold, the related inhomogeneous broadening of the modes is observed, while in the condensed regime discrete energy modes associated to monolayer thickness fluctuations can be resolved, (supplementary A). Figure 1-e shows the first five accessible Lower Polariton Branches (LPBs) along the thickness gradient. The Rabi splitting is about 210 meV for a $\lambda$ cavity at 300 K.

Figure 2 presents the µPL experiments performed at 300 K (see methods). For the investigated 2.5 $\lambda$ cavity region, a single LPB is present within the depicted energy range. The exciton-photon detuning is close to zero, allowing to achieve the lowest polariton laser threshold at this temperature. The angle-integrated PL spectra are shown in figure 2-a. A clear threshold behavior corresponding to an exponential rise of the intensity by three orders of magnitude (fig 2-b) is demonstrated. The linewidth of the emission (fig 2-b) falls down to 0.4 meV above threshold, which is a clear signature of the onset of temporal coherence. This linewidth is 12 times sharper than in Ref. 29. Angular and energy-resolved emission at three different pumping powers are presented on figure 2 e,f,g. The condensate appears centered on k=0, blue shifted by about 4 meV above the polariton dispersion. The blue shift value is only 5 % of the energy difference between the polariton and the bare photonic mode, in sharp contrast with the 50% value shown in Ref. 29. The threshold value itself is about one order of magnitude smaller than in our previous experiments [16,28] performed using an epitaxial cavity. The angular aperture of the main emission line is here about 13° which is narrow compared with state-of-the-art ZnO nano-lasers[11]. The emission area is of the order of 6 µm (fig 2-d) which contrasts with condensation reported in epitaxial structures, where the condensates are typically localized in sub-micrometer traps. The blue shift and angular broadening can be explained by the interaction of the condensate with the cloud of uncondensed excitons, which creates a local repulsive potential expanding the condensate both in real and reciprocal space[25].

The polariton laser threshold versus temperature (4-300 K) and LPB energy is shown in Figure 3-a. The corresponding theoretical values obtained by the numerical solution of semi-classical Boltzmann equations (Supplementary B) are shown in figure 3-b. Similarly to phase diagrams measured in other material systems[20,21,22], for a given temperature the threshold first decreases versus detuning, passes through a minimum and then increases. The optimal LPB offset is determined by the trade-off between thermodynamics and kinetics. This offset increases versus temperature because the kinetics becomes more favorable, contrary to the thermodynamic limit, which degrades. In contrast with the observations in GaN based cavities[21] and for the same temperature range, the optimal detuning (which is directly related with the LPB offset: see Supplementary B) of our ZnO based structure is positive, as theoretically expected[15]. Positive detuning means that the exciton fraction exceeds 50 %. This behavior is due to the large Rabi splitting value, which induces a very deep energy minimum at the bottom of the LPB while keeping a large density of final states and fast relaxation kinetics. Another characteristic feature is the global increase of the threshold for temperatures rising above 100 K. We expect this rise to be induced by the shortening of the non radiative life time versus temperature, which is deliberately not accounted for in the simulations. Interestingly, the polariton lasing threshold versus detuning at low temperatures shows a double-dip structure (fig 3.a-inset). The existence of a second optimum is induced by efficient polariton scattering directly from the reservoir to the LPB, as we recently suggested[28]. Such LO-phonon assisted relaxation (although much less efficient) was already evidenced in CdTe-based structures[20]. At intermediate temperatures (100 K), the two dips merge to form a single minimum, and this is where we obtain the overall lowest pumping threshold. At higher temperature, the optimal detuning value continues to increase, reaching a value compatible with a relaxation assisted by the emission of 2 LO Phonons. This is well reflected in the experimental optimal LPB offset versus temperature (Fig. 3-c). The efficiency of this process allows polariton relaxation towards the LPB despite its large offset. This mechanism, combined with the large density of final states induced by the large Rabi splitting (3 to 5 times larger than in GaN), provides fast scattering rates and is responsible for the low threshold, as supported by the small blue shift values (2-10 meV).

Outstandingly, at low temperature condensation is observed up to an LPB offset of only 25 meV from the bare exciton, which corresponds to a record excitonic fraction of 96 % for a polariton condensate. On the opposite side of the diagram, at 300 K, this fraction falls down to 12 %. For these two situations, the effective masses of the condensed polaritons, which are proportional to their photon fraction, are changing by a factor 20 and this is reflected in the spectral and angular emission pattern below and above threshold, as shown in figures 3-d and e. Below threshold, the exciton-like dispersion is spectrally narrow, being very weakly affected by the photonic inhomogeneous broadening induced by the thickness gradient. Above threshold, the condensate appears well above the bare dispersion and wide in k-space, as it is strongly perturbed by the interaction with the uncondensed excitonic reservoir. In the strongly photonic case, below threshold the dispersion demonstrates large inhomogeneous broadening, with several neighbouring modes being visible. On the other hand, above threshold the photonic condensate clearly shows up within the bare dispersion, less affected by the local exciton reservoir localized below the 2 micro-meter pumping laser. Therefore, this system offers a unique opportunity to study properties of polariton condensates on an unprecedented range of parameters, from low to high temperatures and from strongly excitonic to strongly photonic contents, as further illustrated below.

A remarkable regime can be achieved in the thicker region of the sample, where several LPBs are simultaneously present. Depending on the above mentioned trade-off between the optimal thermodynamic (larger LPB offset) and kinetic (smaller LPB offset) requirements, polaritons "choose" the optimal branch to condense. This is illustrated in figure 4-a, where the energies of three LPBs are shown versus the position on the sample. The modes in which polariton condensation takes place are connected with a solid line. As the LPB offset increases, lasing is continuously observed from the same mode. When the mode becomes too much photonic-like, the system switches and lasing takes place on the next mode, which is closer to the excitons and more kinetically favored. Furthermore, for a given LPB offset (comparable to point 7) the switching can be controlled by the pumping, as shown on figures 4-b,c,d. Figure 4-b (below threshold) shows three polariton modes with very different excitonic contents. Figure 4-c (at threshold) demonstrates lasing from the intermediate branch LPB1 (7 % photon). The relaxation toward this branch is indeed kinetically easier, whereas particles cannot reach the LPB0, which would be thermodynamically favorable. A further increase of pumping enhances exciton-exciton scattering processes and the condensate is then able to reach a lower energy state, at the bottom of LPB0, which is 27 % photonic (figure 4-c). This process can be used to realize a tunable laser and is also very promising for room-temperature optical switches. Interestingly, the photon fraction of the condensate increases by a factor 4 by passing from one branch (LPB1) to another (LPB0). The typical potential seen by the polaritons in both cases is shown in figure 4e. It is composed of a Gaussian potential induced by the excitonic reservoir, which acts on the excitonic part of the polaritons, and a staircase potential acting on the photonic part only. The height of a step corresponds to 1 monolayer variation of the cavity thickness. A simulation (see Supplementary D) based on the numerical solution of the Schrodinger equation using the above-mentioned potential reproduces very well the observed features (fig 4f) and gives further insight into the behavior of the condensate in the current cavity. The k-distribution of the condensate occurring on the LPB1 shows a wide homogeneous peak due to the interaction with the excitonic reservoir. On the other hand, the more photonic condensate shows a series of discrete well-defined sharp peaks up to large wave vectors values. This is due to the acceleration of the condensed polaritons by the staircase potential. Under this situation, the condensate propagates along the thickness gradient, away from the pumping area, and crosses the successive polariton modes associated with the discrete values of the cavity thickness (see Supplementary D).

We report on the fabrication of a fully hybrid ZnO cavity combining high material quality and high Q-factor. Low threshold polariton lasing is obtained from 4 to 300 K with a well collimated laser beam. A phase diagram is established and the main advantages of the use of ZnO, namely the large light-matter coupling and the efficient LO-phonon assisted relaxation, have been demonstrated. Unique features of the polariton physics have been evidenced, such as the creation of a polariton condensate with 96 % exciton fraction and switching induced by mode competition. We also would like to point out that the residual optical absorption in the employed DBRs is relatively low. There are essentially no limitations to increase the number of DBR pairs which could allow to achieve Q factors one order of magnitude higher in the nearest future. Beyond the applied interest, these new hybrid structures open the very interesting perspective to perform propagation experiments at room temperature and to study the temperature and exciton fraction dependence of the superfluid behavior and topological defects in polariton fluids.

## **Acknowledgments**

We thank J.Y. Duboz and B. Gil for their support. We acknowledge funding from the FP7 ITN Clermont4 (235114) and Spin-Optronics (237252), and IRSES Polaphen (246912).

## **Methods**

The micro-photoluminescence experiments reported in this work have been performed using two different excitation techniques. Data shown in the figures 2 and 3-a,c,e, have been measured under a non-resonant and quasi-continuous excitation with a 266 nm Nd:YaG laser (repetition rate 4 kHz and pulse duration 400 ps). Data shown in the figures 3-a-inset, 3-d, 4-a,b,c,d, and Supplementary figures 1, 2, 3 have been recorded under 130 fs pulsed excitation with a 266 nm Ti:Al$_2$O$_3$ laser with a repetition rate of 76 MHz. In both cases, the angle-resolved emission is measured in the far-field by a

Fourier imaging technique within the 0.4 numerical aperture of the microscope objective (spot diameter ~2 µm).

## References

1 Alferov Z. I., Nobel lecture: The double heterostructure concept and its applications in physics, electronics, and technology, *Reviews of modern physics* **73**, 767, (2001).

2 Imamoglu A. and Ram R.J., Quantum dynamics of exciton lasers, *Phys. Lett. A*, **214**, 193, (1996).

3 Kavokin A., Malpuech G., *Cavity Polaritons,* Edited by V.M. Agranovich, *Elsevier North Holland* (2003).

4 Kazprzak J. et al. Bose-Einstein condensation of exciton polaritons, *Nature*, **443**, 409 (2006).

5 Balilli R., et al. Bose Einstein Condensation of microcavity polaritons in a trap, *Science* **316**, 1007, (2007).

6 Christopoulos S. et al. Room-Temperature Polariton Lasing in Semiconductor Microcavities, *Phys. Rev. Lett.* **98**, 126405, (2007).

7 Baumberg J.J. et al. Spontaneous Polarization Buildup in a Room-Temperature Polariton Laser, *Phys. Rev. Lett*. **101**, 136409 (2008).

8 Christmann G.et al., Room temperature polariton lasing in a GaN/AlGaN multiple quantum well microcavity, *Appl. Phys. Lett*. **93**, 051102, (2008).

9 Tang Z.K. et al. Room-temperature ultraviolet laser emission from self-assembled ZnO microcrystallite thin films, *Appl. Phys. Lett.* **72**, 3270 (1998).

10 Huang M.H et al., Room-Temperature Ultraviolet Nanowire Nanolasers, *Science*, 292, 1897, (2001).

11 Chu Sheng et al. Electrically pumped waveguide lasing from ZnO nanowires, *Nature Nanotechnology*, **6**, 506, (2011).

12 K.H. Liang et al., Directional and controllable edge-emitting ZnO ultraviolet random laser diodes, *Appl. Phys. Lett.*, **96**, 101116, (2010).

13 Kalusniak S., Sadofev S., Halm S., Henneberger F., Vertical cavity surface emitting laser action of an all monolithic ZnO-based microcavity, *Appl. Phys. Lett.* **98**, 011101 (2011).

14 Zamfirescu M, Kavokin A, Gil B, Malpuech G, Kaliteevski M, ZnO as a material mostly adapted for the realization of room-temperature polariton lasers*, Phys. Rev. B* **65**, 161205, (2002).

15 Johne R., Solnyshkov D., Malpuech G., Theory of exciton-polariton lasing at room temperature in a ZnO microcavity*, Appl. Phys. Lett*. **93**, 211105 (2008).

16 Guillet T. et al., Polariton Lasing In a Hybrid Bulk ZnO Microcavity, *Appl. Phys. Lett.*, **99**, 161104, 2011.

17 Franke H., Sturm C., Schmidt-Grund R., Wagner G., Grundmann M., Ballistic propagation of exciton-polariton condensates in a ZnO microcavity, *New J. Phys.*, **14**, 013037, (2012).

18 Trichet A. et al., One-dimensional ZnO exciton polaritons with negligible thermal broadening at room temperature, *Phys. Rev. B (R)* **83**, 041302, (2011).

19 Xie W. et al., Room-Temperature Polariton Parametric Scattering Driven by a One-Dimensional Polariton Condensate, *Phys. Rev. Lett*. **108**, 166401 (2012).

20 Kazprzak J., Solnyshkov D.D., André R., Le Si Dang, Malpuech G., Formation of an exciton polariton condensate: thermodynamic versus kinetic regimes, *Phys. Rev. Lett.* **101,** 146404, (2008).

21 Levrat J. et al., Condensation phase diagram of cavity polaritons in GaN-based microcavities: Experiment and theory**,** *Phys. Rev. B* **81**, 125305 (2010).

22 Wertz E. et al., Spontaneous formation of a polariton condensate in a planar GaAs microcavity, *Appl. Phys. Lett.* **95**, 051108 (2009).

23 Amo A. et al., Collective fluid dynamics of a polariton condensate in a semiconductor microcavity *Nature*, **457**, 291 (2009).

24 Lagoudakis, K. G., et al., Observation of Half-Quantum Vortices in an Exciton-Polariton Condensate. *Science* **326**, 974-976 (2009).

25 Wertz, E. et al. J. Spontaneous formation and optical manipulation of extended polariton condensates, *Nature Phys.* **6**, 860-864 (2010).

26 Sich M. et al*,* Observation of bright polariton solitons in a semiconductor microcavity, *Nature Phot.* **6**, 50 (2012).

27 Kéna-Cohen, S. & Forrest, S. R. Room-temperature polariton lasing in an organic single-crystal microcavity, *Nature Photonics* **4**, 371 - 375 (2010).

28 Orosz L., LO-phonon-assisted polariton lasing in a ZnO-based microcavity*, Phys. Rev. B* **85**, 121201 (2012).

29 Lu Tien-Chang et al., Room temperature polariton lasing vs. photon lasing in a ZnO-based hybrid microcavity, *Optics Express*, **20**, 5530, (2012).

30 Orosz L. et al., Fabrication and Optical Properties of a Fully-Hybrid Epitaxial ZnO Based Microcavity in the Strong-Coupling Regime, *Appl. Phys. Express*, **4**, 072001, (2011).

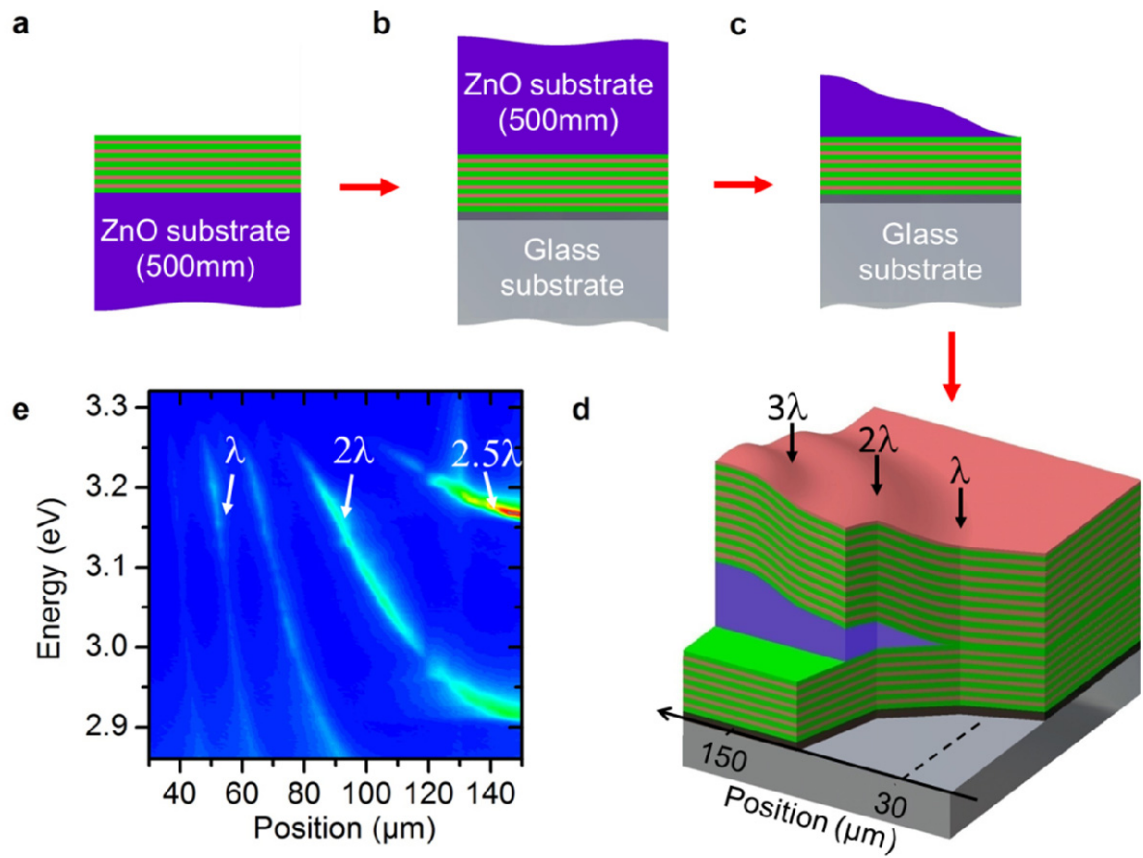

**Figure 1: Fabrication of the fully-hybrid bulk ZnO microcavity.**

**a**, A single-crystal c-ZnO substrate is covered by a $HfO_2/SiO_2$ DBR. **b**, The whole DBR/ZnO substrate is flipped upside-down and transferred to a glass substrate. **c**, The ZnO substrate is backside etched/polished down to a thickness of several tens to hundreds nanometers. **d**, A $HfO_2/SiO_2$ DBR is deposited on top of the processed ZnO substrate. **e**, Emission intensity below threshold measured along the thickness gradient and showing the first five polariton modes (cavity thickness from $\lambda/2$ to $5\lambda/2$).

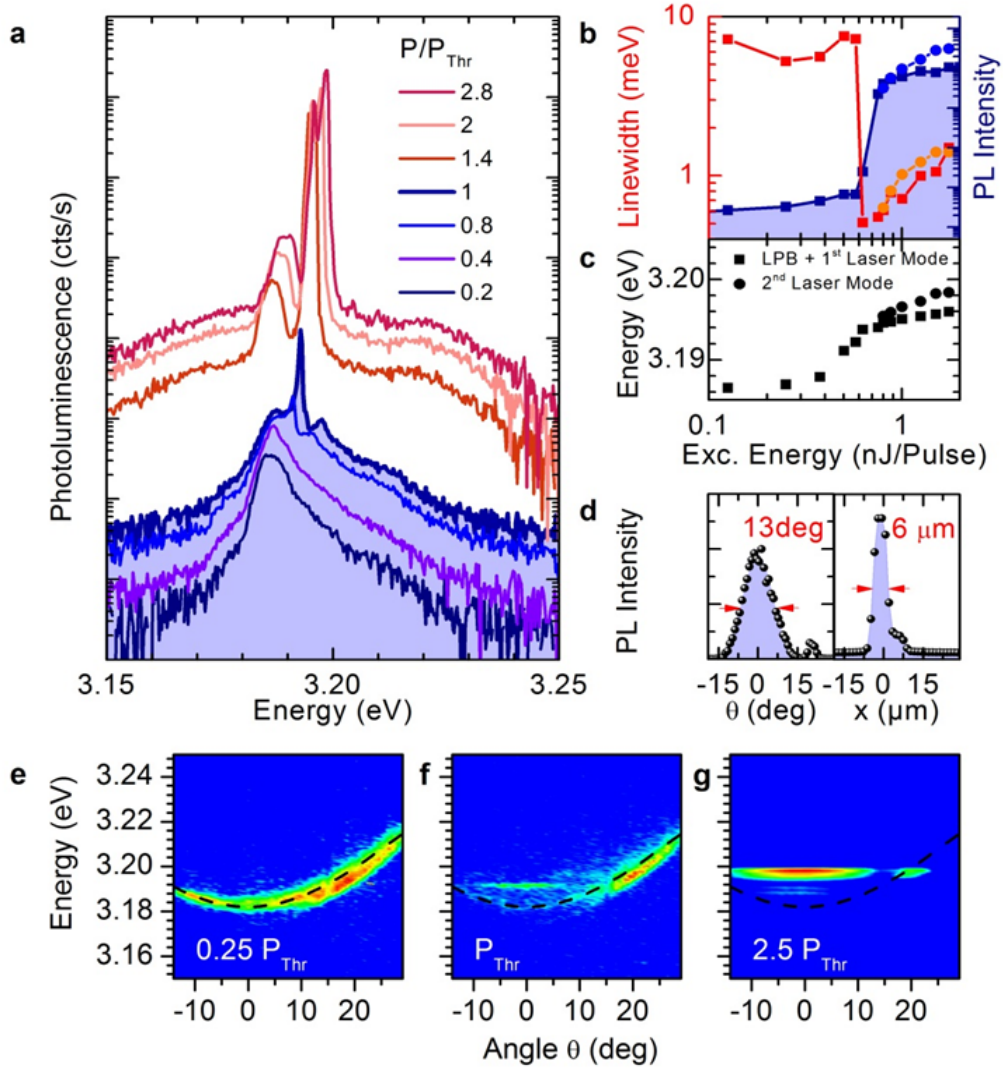

**Figure 2 : Polariton lasing at 300 K**

(Thickness of the cavity: 2.5 λ, δ=-8 meV, for a Rabi splitting of 230 meV, i.e a photonic fraction of 52% for the LPB)

**a,** Power dependent series of angle integrated micro-PL spectra. The polariton lasing threshold ($P_{thr}$ = 0.6 nJ/pulse) corresponds to the thick blue spectrum. The excitation power density relatively to the threshold power is indicated for each spectrum. **b,c,** Integrated intensity, linewidth and energy of the LPB transition and the first laser mode (square) and the second laser mode (circle) as a function of the excitation power density. **d,** Spatial and angular distribution of the emission above threshold at a similar detuning. **e,f,g,** Fourier space images of the polariton emission as a function of excitation power. The black dashed lines represent the result of the two-level coupled oscillator model.

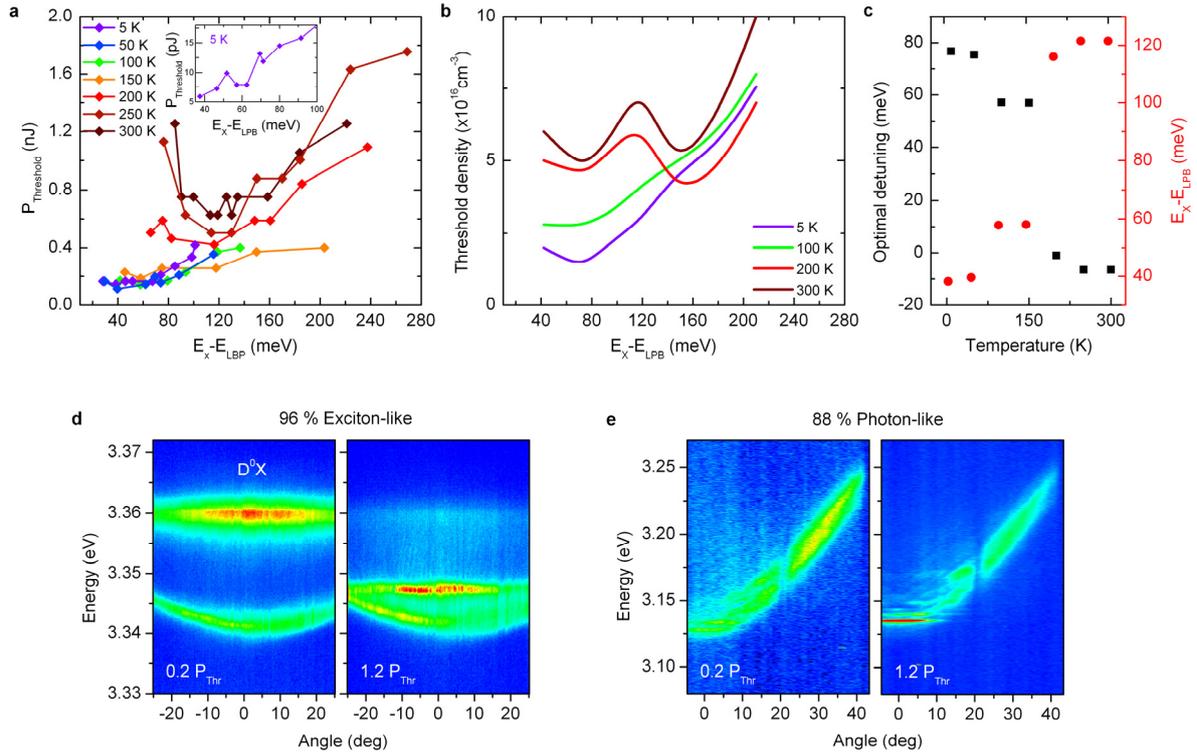

**Figure 3: Polariton laser phase diagram: from Exciton-like to Photon-like condensates**

**a,** Polariton laser threshold measured under ns excitation, from 5 K to 300 K, versus the energy offset between the exciton energy and the LPB. Caption: Double dip dependence of the threshold against the offset, better revealed by fs excitation **b,** Polariton laser threshold density calculated for the same range of parameters as in **a**, using the semi-classical Boltzmann equation detailed in the supplementary Suppl. C. **c,** Temperature dependence of the optimal offset and corresponding optimal exciton-photon detuning which minimizes the polariton laser threshold. **d,** Fourier space images of the polariton emission with 96 % exciton fraction (4% photon fraction), below and above the condensation threshold. **e,** Fourier space images of the polariton emission with 12 % exciton fraction (88% photon fraction), below and above the condensation threshold.

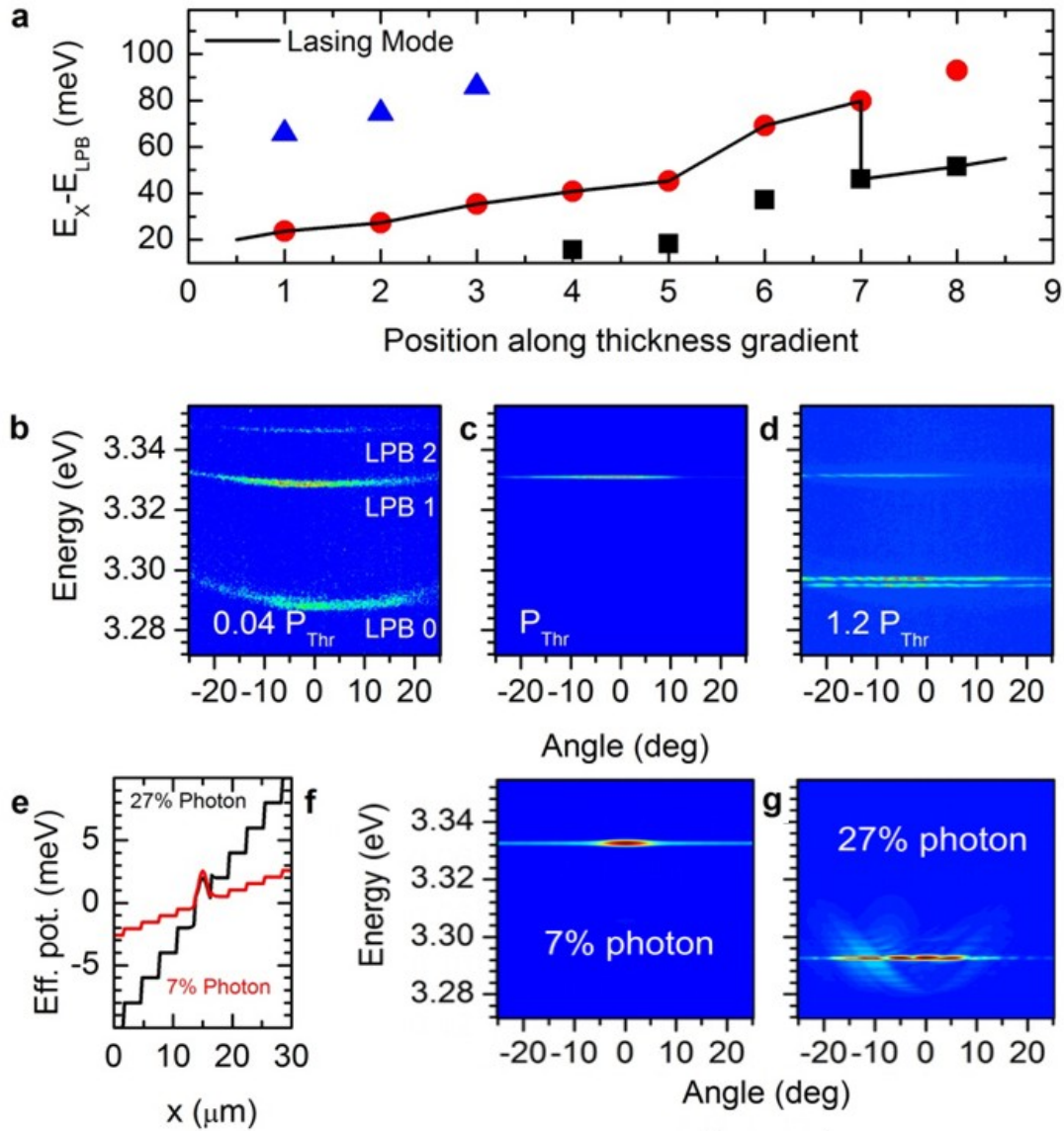

**Figure 4: Mode switching: From exciton-like to photon-like condensate.**
**a,** Polariton mode offset versus the position along the thickness gradient of the cavity. Three LPB modes (plotted as triangle, circular and square symbols, respectively) are simultaneously seen in this region. The solid line binds the modes on which the polariton lasing occurs. At point 7, the lasing threshold difference between the lasing branches is less than 25%. The data are taken under fs excitation. **b,c,d,** Fourier space images illustrating mode switching as a function of pumping intensity (situation similar to point 7 in panel **a**). Three LPBs, labelled from 2 to 0 with increasing offset, are seen with 99%, 93%, and 73% exciton fraction, respectively. **b,** Image below the lasing threshold. **c**, Image taken at threshold. Lasing occurs on LPB1. **c,** At only 1.2 Pth, the polariton condensate switches on the thermodynamically favourable mode LPB0. **e,** Effective potential seen by the polaritons on LPB1 and LPB0 respectively. **f**,**g,** Calculated emission of the polariton condensate created under the 2 μm pumping area and propagating in the potentials shown on the panel **e** (see Suppl. D for details).

# ZnO-based polariton laser operating at room temperature: From Excitonic to Photonic condensate

Feng Li[1,2], Laurent Orosz[3], Olfa Kamoun[4], Sophie Bouchoule[5], Christelle Brimont[4] Pierre Disseix[3], Thierry Guillet[4], Xavier Lafosse[5], Joel Leymarie[3], Meletis Mexis[4], Martine Mihailovic[3], François Réveret[3], Dmitry Solnyshkov[3], Jesus Zuniga-Perez[1], Guillaume Malpuech[3]

[1] CRHEA-CNRS, Rue Bernard Gregory, 06560 Valbonne, France

[2] Université de Nice Sophia-Antipolis, 06103 Nice, FRANCE

[3] Institut Pascal, PHOTON-N2, Clermont Université, CNRS and Université Blaise Pascal, 24 Avenue des Landais, 63177 Aubière cedex, FRANCE

[4] Université de Montpellier 2, Laboratoire Charles Coulomb, UMR 5221, 34095 Montpellier, FRANCE

[5] LPN-CNRS, Route de Nozay, 91460 Marcoussis, FRANCE

**A Demonstration of the high quality factor**

As presented in the first part of the article, special attention was devoted to the elaboration of high optical quality ZnO combined with a large cavity quality factor. Indeed, a long photon lifetime is required to reduce the polariton laser threshold. The latest improvement of the Bragg mirror interfaces allows to reach a record quality factor of 2,600, rarely observed for UV planar cavities. The Supplementary Figure 1a presents the corresponding emission spectrum for $\theta$ = 0 deg extracted from a µPL measurement. The energy of the polaritonic mode is far below the exciton energy (>1LO) and the full width at half maximum (FWHM) of the peak (1.2 meV) is only influenced by the low thermal excitonic broadening [1]. A two-levels coupled oscillator model is used to estimate both the photonic and the excitonic part of the polariton. Due to the large Rabi splitting, even if the energy of the polariton at normal incidence is 130 meV below the exciton, the polaritonic mode is 52% photon-like. It means that the polaritonic linewidth is broadened by its thermal excitonic fraction (2 meV), consequently the off-resonance Q-factor is higher than 2,600. This measured value is however affected by the photonic inhomogeneous broadening induced by the large thickness gradient in the structure. This specificity is clearly evidenced in the condensed regime, where the condensate is accelerated by the cavity thickness gradient and expands in space, exciting a whole series of individual polariton modes corresponding to different thicknesses of the ZnO layer. Just at threshold, when the condensed fraction itself is not yet too strong, we can very well resolve these individual dispersions corresponding to different locations on the sample (Supplementary Figure 1b). The individual modes are not affected anymore by the inhomogeneous broadening and are quite sharp (typically 0.5-0.8 meV). These values are compatible with Q factors of the order of 4,000 and correspond to particle life times of the order of 1 ps, which is similar to the life time value in the CdTe cavity where polariton BEC was demonstrated [2].

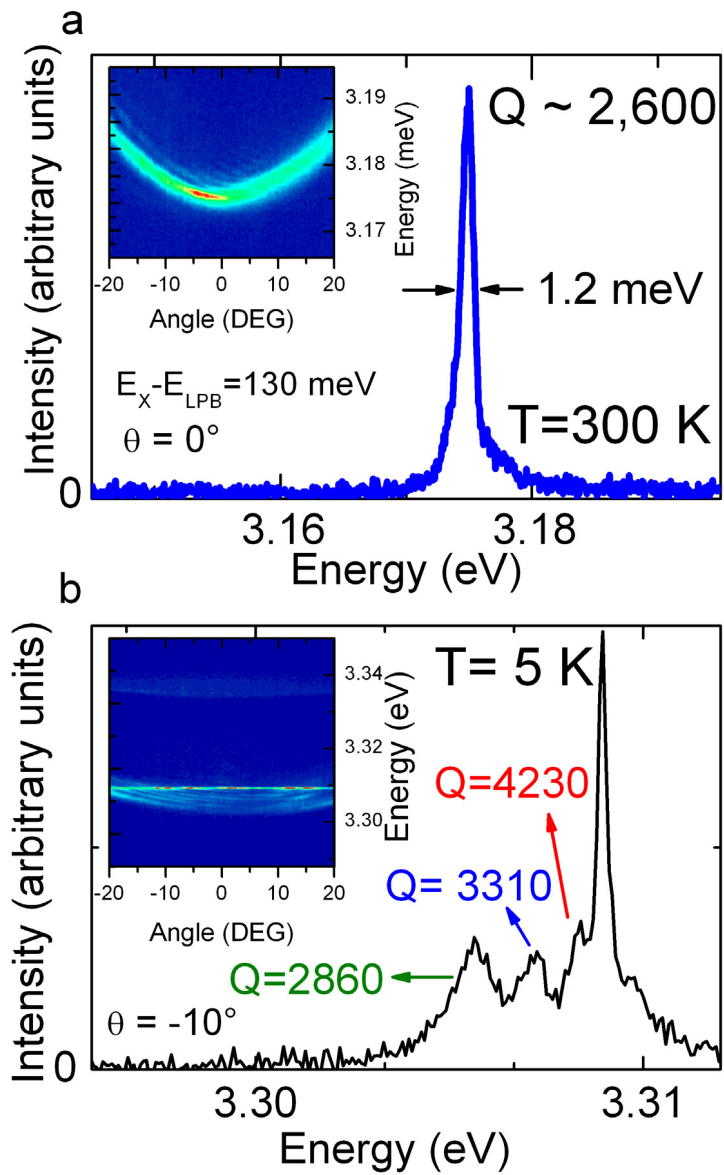

**Supplementary Figure 1 : Quality factor measurement**. **a**, Micro-photoluminescence emission spectrum taken at 300 K, below the condensation threshold. The cavity thickness is 2.5 lambda. The width at half maximum of the emission line is 1.2 meV. The inset shows the Fourier image of emission. **b**. Emission spectrum taken at -10° on another point of the sample at 5 K. The width of the individual modes can be extracted and are as small as 0.8 meV, corresponding to Q-factors values of the order of 4,000. The inset shows Fourier space image of the polariton emission taken at the condensation threshold. The presence of discrete individual modes is clearly visible.

Also the energy spacing between the modes is here typically 1.2 meV. The corresponding energy difference between the photonic modes associated with the polaritons is therefore about 5 meV.

This value is consistent with the change of the cavity width by 0.15 %, corresponding to the change of the ZnO thickness by steps of one monolayer.

## B Estimation of the Rabi splitting

As shown in Figure 1 of the paper, the active layer presents a significant thickness gradient. Therefore, the Rabi splitting varies from place to place preventing a direct determination of the detuning.

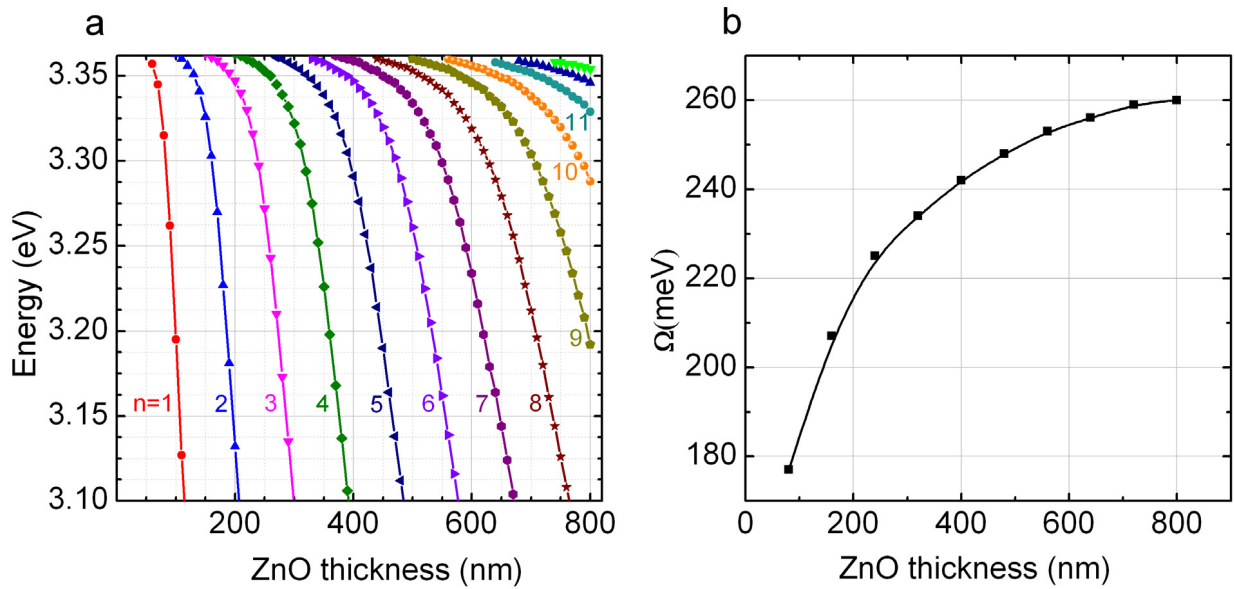

**Supplementary Figure 2 : Rabi Splitting determination**
**a** Evolution of the energy of the cavity modes as a function of ZnO thickness at 70 K, for normal incidence. The n index allows to identify each mode. **b** Calculated Rabi splitting of the cavity versus the ZnO thickness.

The use of the transfer matrix formalism allows the calculation of the reflectivity spectrum of the sample for various thicknesses. The employed ZnO excitonic parameters have been previously deduced from a careful analysis of continuous-wave reflectivity experiments combined with measurements of autocorrelation spectroscopy.

For each thickness, the energy of the optical modes are then determined by comparing the PL measurements giving access to energy offset (energy difference between the LPB and the Exciton energy) with, the curves shown on the Supplementary Figure 2a (T=70 K). It has to be noted that the number of allowed modes increases with the increase of the ZnO thickness.

The Rabi splitting and the exciton-photon detuning are deduced from the calculations through the determination of the minimum energy difference between the lower and upper polariton branches. The Rabi splitting is revealed by removing the band-to-band absorption in the calculations. The evolution of the Rabi splitting as a function of the ZnO thickness at 70 K is displayed on Supplementary Figure 2b. The exciton-photon detuning is defined as the energy difference between the bare exciton and bare photon modes. One therefore understand, that the Rabi splitting, exciton-photon detuning, and exciton-photon fractions are not directly measured, but extracted from the data by their comparison with the result of the transfer matrix method, and two oscillator model.

## C Boltzmann equation formalism

Here we describe the simulations of polariton relaxation shown in Figure 3 of the main text. The simulations are based on numerical solution of semi-classical Boltzmann equations, according to the approach developed and described mainly in Refs. 3 and 14 of the main text. The main equation describing the time dynamics of the polariton distribution function $n_{\mathbf{k},i}$ depending on the wavevector $\mathbf{k}$ and branch number $i$ reads:

$$\frac{dn_{\mathbf{k},i}}{dt} = P_{\mathbf{k},i} - \Gamma_{\mathbf{k},i} n_{\mathbf{k},i} - n_{\mathbf{k},i} \sum_{\mathbf{k}',i'} W_{\mathbf{k},i \to \mathbf{k}',i'} \left( n_{\mathbf{k}',i'} + 1 \right) + \left( n_{\mathbf{k},i} + 1 \right) \sum_{\mathbf{k}',i'} W_{\mathbf{k}',i' \to \mathbf{k},i} n_{\mathbf{k}',i'} ,$$

where the terms $\left( n_{\mathbf{k},i} + 1 \right)$ and $\left( n_{\mathbf{k}',i'} + 1 \right)$ account for bosonic stimulation, $P_{\mathbf{k},i}$ are the pumping terms, $\Gamma_{\mathbf{k},i}$ describes the particle decay. In the calculation of the dispersion $E_i(\mathbf{k})$ and of the scattering rates $W_{\mathbf{k},i \to \mathbf{k}',i'}$, we account for 2 lower polariton branches with 2D densities of states and an excitonic reservoir with a 3D density of states, since the active layer of the cavity is bulk. We assume that the pump laser creates excitons in the reservoir with an effective temperature given by the LO-phonon energy, and the relaxation towards the polariton branches occurs through polariton-phonon and polariton-polariton interactions, described by the scattering rates $W_{\mathbf{k},i \to \mathbf{k}',i'}$ calculated with the Fermi's golden rule. Both acoustic and optical phonons are taken into account. Cylindrical symmetry of the polariton distribution in the reciprocal space is assumed, and the scattering processes between different wavevectors are properly accounted for. The parameters used in the simulation are LO-phonon energy $E_{LO}$=72 meV, exciton Bohr radius $a_B$=18 Å, effective electron mass $m_e$=0.24$m$, and effective hole mass $m_h$=0.98$m_0$, where $m_0$ is the free electron mass, deformation potential D =15 eV, exciton lifetime $\tau_{ex}$ =50 ps, cavity quality factor Q=2500 and inhomogeneous broadening of 1 meV.

The scattering rates are calculated using the Fermi's golden rule:

$$W_{\mathbf{k},i \to \mathbf{k}',i'} = \frac{2\pi}{\hbar} \left| M_{\mathbf{k},i \to \mathbf{k}',i'} \right|^2 D\left( E(\mathbf{k}',i') \right),$$

where $M_{\mathbf{k},i \to \mathbf{k}',i'}$ is the matrix element of interaction. These scattering rates are therefore directly proportional to the final density of states $D\left( E(\mathbf{k}',i') \right)$, which has a crucial impact on the system's behavior. Indeed, as already pointed out in ref 14 of the main text, the depth of the polariton trap ($E_{LPB}$-$E_X$) can be increased either by going to negative detunings or by increasing the Rabi splitting. The efficiency of the relaxation is very different in the two cases, because negative detuning means strong photonic fraction, light effective mass, and, therefore, a low density of states, while positive detuning with strong Rabi splitting gives exactly the opposite, that is, a large density of states.

The squared matrix elements of the polariton-phonon and polariton-polariton interactions are, of course, directly proportional to the excitonic fractions $x_k$ of the polariton states involved, since only excitonic part of the polariton can interact with the surrounding media. For example, the polariton-phonon interaction can be written as:

$$\left|M_{k\to k'}(\mathbf{q})\right| = \left|\left\langle \psi_k^{pol} \left| H_{pol-ph}^q \right| \psi_{k'}^{pol} \right\rangle\right| = \sqrt{x_k x_{k'}} \left|\left\langle \psi_k^{exc} \left| H_{exc-ph}^q \right| \psi_{k'}^{exc} \right\rangle\right|$$

Therefore, the efficiency of all relaxation mechanisms is increased at larger excitonic fractions and larger Rabi splittings, through both the density of states and the excitonic fraction dependencies. The Figure 3 of the main text is derived considering the steady-state solution of the rate equations under a continuous and spatially homogeneous pumping of the exciton reservoir.

**D Modeling and evidence of the polariton gas spatial expansion in the condensed regime.**

Here, we describe the model used to calculate the Figures 4f and 4g of the main text, which demonstrate the different behavior of the photon-like, and exciton-like polariton condensates. We also show additional experimental results, which strongly suggest that the reciprocal space pattern of the emission can be attributed to a spatial expansion of the photon-like polariton condensate.

The polariton system is described by two coupled Gross-Pitaevskii equations, one for the wavefunction of photons $\psi_{ph}$ and one for the wavefunction of excitons $\psi_{ex}$. Here we neglect the non-linear interaction term between condensed particles and the Gross-Pitaevskii equation reduces to the Schrödinger equation:

$$i\hbar \frac{\partial \psi_{ph}}{\partial t} = E_{ph} \psi_{ph} - \frac{\hbar^2}{2m_{ph}} \Delta \psi_{ph} + U_{ph} \psi_{ph} - \frac{i\hbar}{2\tau_{ph}} \psi_{ph} + \frac{\hbar \Omega_R}{2} \psi_{ex} + P_{ph}(t)$$

$$i\hbar \frac{\partial \psi_{ex}}{\partial t} = E_x \psi_x - \frac{\hbar^2}{2m_{ex}} \Delta \psi_{exc} + U_{ex} \psi_{ex} - \frac{i\hbar}{2\tau_{ex}} \psi_{ex} + \frac{\hbar \Omega_R}{2} \psi_{ph}$$

Here $E_{ph}$ is the energy of the cavity mode, $E_{ex}$ is the energy of the excitonic resonance ($E_{ph}$-$E_{ex}$ is the detuning), $m_{ph} = 2 \times 10^{-5} m_0$ is the photonic mass, $m_{exc} \approx m_0$ is the mass of exciton ($m_0$ is the free electron mass), $U_{ph}$ is the staircase potential acting on photons with a step length of 2.5 µm and steps of 5 meV, $U_{ex}$ is the potential due to the interactions with the reservoir, acting on excitons, with a height of 4 meV corresponding to the experiment and width 3 µm, $\tau_{ph}$ is the cavity lifetime corresponding to Q=2500, $\tau_{ex} = 50$ ps is the exciton lifetime, $\hbar\Omega_R = 200$ meV is the Rabi splitting, and P(t) is the pulsed pump, localized in real space (3 µm size), resonant with the lower polariton branch which mimics the condensate creation. By varying the detuning, we have explored both strongly photonic (negative detuning) and strongly excitonic (positive detuning) regimes. The potential shown on the Figure 4f is the one seen by polaritons with photon fractions of 7% and 27% respectively. In the case of more photonic polaritons, the contribution of the staircase potential to the effective potential seen by polaritons is about 4 times higher than in the highly excitonic case. This potential accelerates polaritons in one direction, as previously experimentally evidenced in other cavity system[3]. This leads to the observation of a discrete series of wavevectors, corresponding to several steps of the staircase. In the excitonic case, this potential is washed out, and the only effect observed is the acceleration within the potential created by the excitonic reservoir, leading to the broadening of the maximum of emission in the reciprocal space as shown on the Figure 4f and 4g, in good agreement with the experimental data (Figure 4c,d).

Further evidence of polariton propagation in the condensed regime is shown in the Supplementary Figure 3. Below the condensation threshold, a single dispersion line is visible, with a width which is here of 1.4 meV. This line is inhomogeneously broadened, but the moderate linewidth shows that the spatial expansion of the condensate is also relatively moderate, covering only 1-2 ZnO terraces.

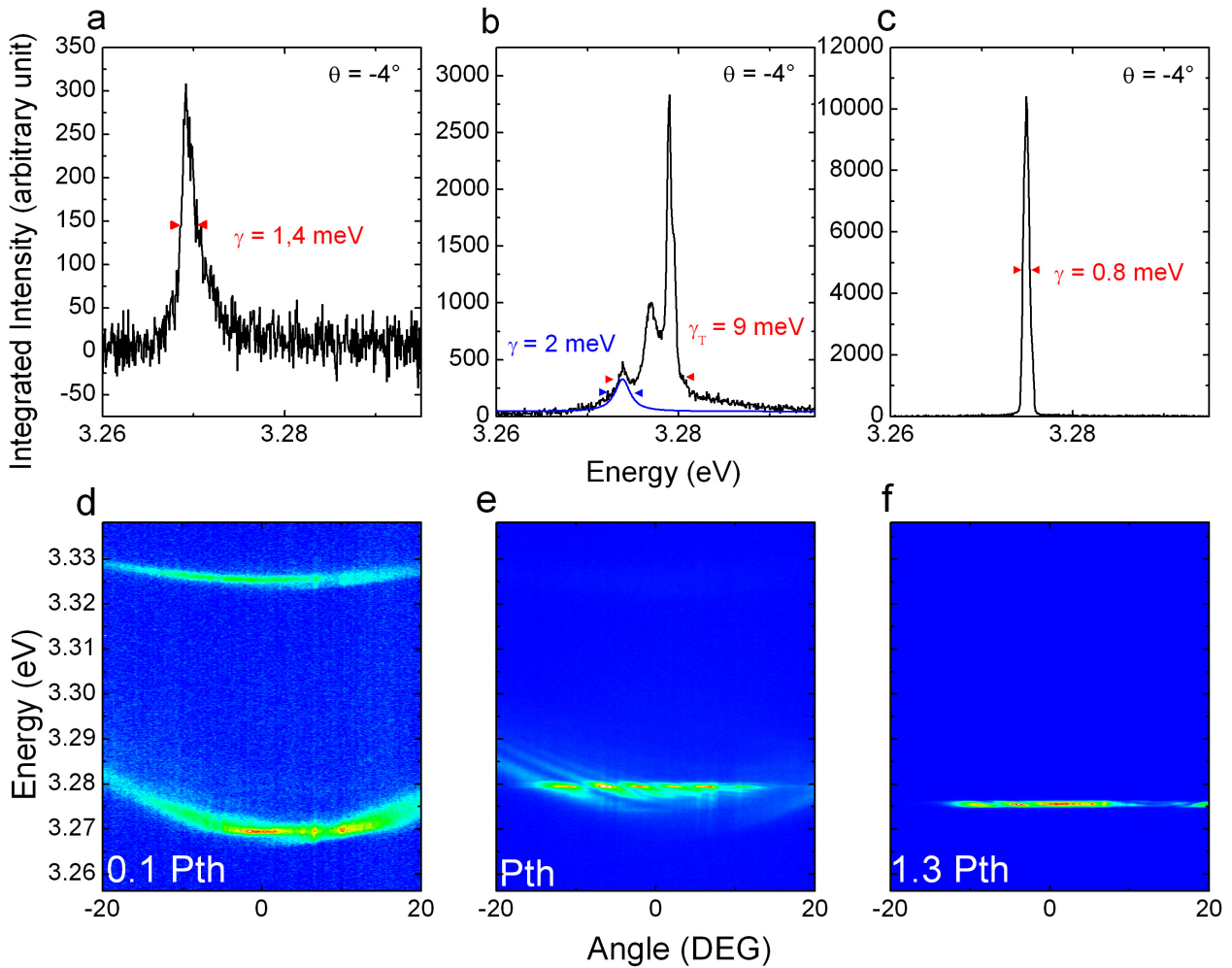

Supplementary Figure 3: **Evidence for the spatial expansion of the polariton cloud in the condensed regime.** (**a,b,c**) Spectra extracted at -4° from Fourier image recorded at T=70K (**d,e,f**) for various excitation intensities. In panels (**a, c**), $\gamma$ indicates the FWHM of the emission. In panel (**b**), $\gamma$ is the FWHM of the emission coming from the lowest energy. $\gamma_T$ is the total width of emission.

Above the condensation threshold, the emission profile is strongly modified. One main sharp line shows up and will completely dominate the emission well above threshold (panels c,f). However, just at threshold, the condensate population is small enough to allow the observation of emission coming from uncondensed polaritons. The corresponding energy width is 9 meV, which is 6 times larger than below threshold and composed by several individual lines. This energy broadening is clearly associated with a spatial expansion of the emitting polariton cloud, which seems to become typically 6 times larger with respect to the uncondensed regime. From the comparison of the panels (e) and (f), it is clear that each bright emission spot in reciprocal space can be associated with one discrete dispersion branch, corresponding to a single ZnO terrace. The condensate is created at k=0 and r=0, and is then accelerated by the thickness gradient, gaining an in-plane wavevector, by moving away from where it was created.

References

1 Trichet A. et al., One-dimensional ZnO exciton polaritons with negligible thermal broadening at room temperature, *Phys. Rev. B (R)* **83**, 041302, (2011).
2 Kazprzak J. et al. Bose-Einstein condensation of exciton polaritons, *Nature*, **443**, 409 (2006).
3 B. Sermage, G. Malpuech, A.V. Kavokin, V. Thierry Mieg, Polariton acceleration in a microcavity wedge, Phys. Rev. B **64**, 081303 (2001).